\documentclass[preprint2]{emulateapj}

\usepackage{psfig}
\usepackage{apjfonts}

\def\approxgt{\ifmmode \rlap{$>$}{}_{{}_{{}_{\textstyle\sim}}} \else%
$\rlap{$>$}{}_{{}_{{}_{\textstyle\sim}}}$\fi} 
\def\approxlt{\ifmmode \rlap{$<$}{}_{{}_{{}_{\textstyle\sim}}} \else%
$\rlap{$<$}{}_{{}_{{}_{\textstyle\sim}}}$\fi}
\def\farcm{\hbox{$.\mkern-4mu^\prime$}}
\def\farcs{\hbox{$.\!\!^{\prime\prime}$}}

\def\arcmin{\hbox{$^\prime$}}
\def\arcsec{\hbox{$^{\prime\prime}$}}

\def\src{1H~1905+000}
\def\flx{erg cm$^{-2}$ s$^{-1}$}
\def\lum{erg s$^{-1}$}

\shorttitle{Upper limit on \src}
\shortauthors{Jonker et al.}

\begin{document}

\title{The cold neutron star in the soft X-ray transient \src}

\author{Peter G.~Jonker\altaffilmark{1,2}}
\affil{SRON, Netherlands Institute for Space Research, Sorbonnelaan 2, 3584~CA, Utrecht, The Netherlands}
\email{p.jonker@sron.nl}
\author{Daniel Steeghs\altaffilmark{3}}
\affil{Harvard--Smithsonian  Center for Astrophysics, 60 Garden Street, Cambridge, MA~02138, Massachusetts,
U.S.A.}
\email{dsteeghs@cfa.harvard.edu}
\author{Deepto Chakrabarty}
\affil{Department of Physics and Kavli Institute for Astrophysics and Space Research, Massachusetts Institute
 of Technology, Cambridge, MA 02139, U.S.A.}
\email{deepto@space.mit.edu}
\and
%\author{E.F.~Brown}
%\affil{Department of Physics and Astronomy, Michigan State University, East Lansing, MI 48824, U.S.A.}
\author{Adrienne M.~Juett}
\affil{Department of Astronomy, University of Virginia, Charlottesville, VA~22903, U.S.A.}
\email{amj3r@astsun.astro.virginia.edu}

\altaffiltext{1}{Harvard--Smithsonian  Center for Astrophysics, 60 Garden Street, Cambridge, MA~02138, Massachusetts,
U.S.A.}
\altaffiltext{2}{Astronomical Institute, Utrecht University, P.O.Box 80000, 3508 TA, Utrecht, The Netherlands}
\altaffiltext{3}{Astronomy \& Astrophysics, Department of Physics, University of Warwick, Coventry
  CV4~7AL, U.K.}

\begin{abstract}

We report on our analysis of 300 ks of {\it Chandra} observations of the neutron
star soft X--ray transient \src\ in quiescence. We do not detect the source down
to a 95\% confidence unabsorbed flux upper limit of $2\times10^{-16}$ \flx\ in
the 0.5--10 keV energy range for an assumed $\Gamma=2$ power law spectral model.
A limit of $1.4\times10^{-16}$ \flx\ is derived if we assume that the spectrum
of \src\ in quiescence is described well with a black body of temperature of 0.2
keV. For the upper limit to the source distance of 10 kpc this yields a 0.5--10
keV luminosity limit of $2.4\times10^{30}$ \lum / $1.7\times10^{30}$ \lum\ for
the abovementioned power law or black body spectrum, respectively. This
luminosity limit is lower than the luminosity of A~0620--00, the weakest black
hole soft X--ray transient in quiescence reported so far. Together with the
uncertainties in relating the mass transfer and mass accretion rates we come to
the conclusion that the claim that there is evidence for the presence of a black
hole event horizon on the basis of a lower quiescent luminosity for black holes
than for neutron stars is unproven. We also briefly discuss the implications of
the low quiescent luminosity of \src\ for the neutron star equation of state.
Using deep Magellan images of the field of \src\ obtained at excellent observing
conditions we do not detect the quiescent counterpart of \src\ at the position
of the outburst optical counterpart down to a magnitude limit of $i'>25.3$. This
can be converted to a limit on the absolute magnitude of the counterpart of
$I>9.6$ which implies that the counterpart can only be a brown or a white dwarf.

\end{abstract}



\section{Introduction}

Low--mass X--ray binaries (LMXBs) are binary systems in which a compact object,
either a neutron star or a black hole, accretes matter from a companion star
that has a mass of typically less than 1~M$_\odot$. Such systems are excellent
test beds for a range of astrophysical questions and probe fundamental physics.
Theories, such as Einstein's Theory of General Relativity, can be tested in the
strong field regime by comparing the observed black hole and neutron star
properties. The presence of an event horizon in the case of a black hole is such
a prediction. Observations of neutron star LMXBs can also help constrain the
equation of state of matter at supra-nuclear densities encountered in the
neutron star core.

Observationally, it was found that in general the luminosity of quiescent black
hole LMXBs is lower than that of quiescent neutron star LMXBs
(e.g.~\citealt{2001ApJ...553L..47G}, \citealt{2002ApJ...570..277K} and
\citealt{2006MNRAS.368.1803J}). However, recent {\it Chandra} and XMM--{\it
Newton} observations of several transient neutron star LMXBs in quiescence have
shown that the luminosities of these sources span a much larger range extending
to lower luminosities than previously thought
(e.g.~\citealt{2005ApJ...635.1233T}, \citealt{2005ApJ...619..492W}). 

The neutron star transient with the lowest quiescent X--ray luminosity is \src.
Using a 25 ks long observation of \src\ with the {\it Chandra} satellite,
\citet{2006MNRAS.368.1803J} did not detect the source in quiescence down to a
luminosity of 1.8$\times10^{31}$ erg s$^{-1}$. This means that the neutron star
luminosity in this source is comparable to that of several quiescent black hole
LMXBs. For a recent introduction on \object{\src}, for a more detailed
introduction on the difference in quiescent luminosity between neutron star and
black hole LMXBs, and for an explanation why these transient neutron stars in
quiescence can provide information on the neutron star equation of state we
refer to \citet{2006MNRAS.368.1803J} (see also \citealt{2004ARA&A..42..169Y} and
\citealt{2006astro.ph.12232H} for the latter subject).

In this Letter, we present our analysis of a long {\it Chandra} and a deep
Magellan observation of \src\ obtained with the aim to detect the source or
provide a stringent limit on the source flux and the source luminosity in
quiescence. 

\section{Observations and analysis}

\subsection{{\it Chandra} X--ray observations}

We have observed 1H~1905+000 with the back--illuminated S3 CCD--chip of the
Advanced CCD Imaging Spectrometer (ACIS) detector on board the {\it Chandra}
satellite. A log of the observations is given in Table~\ref{log}. The data
telemetry mode was set to {\sl very faint} to allow for a thorough background
subtraction. A CCD frame time of 3.04104 s has been used. We have reprocessed
and analysed the data using the {\it CIAO 3.4} software developed by the Chandra
X--ray Center using CALDB version 3.3.0.1, to benefit from the latest
calibrations available early 2007 and to take full advantage of the {\sl very
faint} data mode. In our analysis we have selected events only if their energy
falls in the 0.3--7 keV range in order to reduce the background contamination
that occurs at high energies. The lower cut--off was chosen to avoid calibration
uncertainties below 0.3 keV. Since fluxes and luminosities are commonly provided
in the 0.5--10 keV range, we extrapolate our 0.3--7 keV event rate to model
fluxes over the 0.5--10 keV range in the remainder of this Letter. Data were
excluded for which the 0.3--7 keV background count rate is higher than 0.4
counts s$^{-1}$. The net on--source exposure time is 300.8 ks. 

In the observations presented in Table~\ref{log} one source
(CXOU~J190834.1+001139) is always detected (we will describe the properties of
detected sources unrelated to \src\ or to the presented analysis in a
forthcoming paper). The J2000.0 $\alpha$ and $\delta$ position of that source
was determined in \citet{2006MNRAS.368.1803J} to be $\alpha_{J2000.0}=19{\rm
^h08^m34^s.108}\pm$0\farcs033, $\delta_{J2000.0}=
+00^\circ11'39$\farcs01$\pm$0\farcs032. We use this source to apply a boresight
correction to each of the observations separately (see Table~\ref{log}) before
combining them. In addition we applied a boresight correction of ($\alpha$,
$\delta$)=0\farcs090$\pm$0\farcs013,-0\farcs208$\pm$0\farcs014 to the final
combined image such that the coordinates for CXOU~J190834.1+001139 measured on
the combined image are, within errors, consistent with the optical coordinates.
The latter boresight correction accounts for the limited accuracy with which the
source position can be determined in the individual observations.

\begin{deluxetable}{cccc}
\tabletypesize{\scriptsize}
\tablecaption{Log of the {\it Chandra} observations  \label{log}}
\tablewidth{0pt}
\tablehead{
\colhead{ID} & \colhead{T$_0$} & \colhead{Exposure\tablenotemark{a}}
& \colhead{Boresight corr.} \\
\colhead{} & \colhead{({\sl UTC})} & \colhead{(ks)} 
& \colhead{($\alpha$, $\delta$)} }
\startdata
5549 &  Febr.~25, 2005  & 24.84  &  0\farcs128$\pm$0\farcs044,-0\farcs074$\pm$0\farcs041  \\ 
6649 &  Sept.~17, 2006  & 151.92&  0\farcs314$\pm$0\farcs018,0\farcs611$\pm$0\farcs019  \\ 
6650 &  Sept.~20, 2006  & 40.63 &  -1\farcs07$\pm$0\farcs030,0\farcs047$\pm$0\farcs039  \\ 
8261 &  Sept.~17, 2006  & 41.61 &  -1\farcs08$\pm$0\farcs029,0\farcs539$\pm$0\farcs033  \\ 
8262 &  Sept.~22, 2006  & 38.65 &  -0\farcs789$\pm$0\farcs025,0\farcs392$\pm$0\farcs030  \\ 
8283 &  Sept.~19, 2006  & 3.15  &  -0\farcs820$\pm$0\farcs066,0\farcs758$\pm$0\farcs087  \\ 
\enddata
\tablenotetext{a}{After taking into account the CCD read--out time and after filtering for flares.}
\end{deluxetable}

\begin{figure}
\epsscale{.80}
\plotone{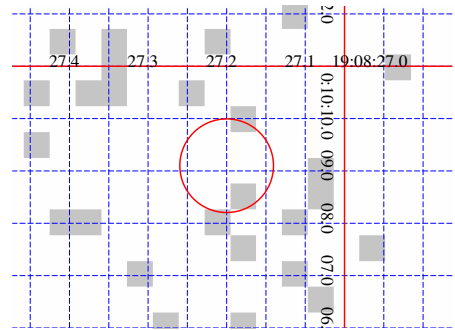}
\caption{Zoom--in on the position of the neutron star SXT \src\ in the 
0.3--7 keV 300 ks {\it Chandra} ACIS image. The circle has a radius of 0\farcs91.
\label{zoom1905}}
\end{figure}

The outburst position of \src\ is $\alpha_{J2000.0}=19{\rm
^h08^m27^s.200}\pm$0\farcs084, $\delta_{J2000.0}=
+00^\circ10'09$\farcs10$\pm$0\farcs087 (\citealt{2006MNRAS.368.1803J}). The
largest off--axis angle that the source position was observed at was
$\theta=$0\farcm54 (see Table~\ref{log}). Using the analytical expression for the
90\% encircled energy radius, R$_{90}$ (in \arcsec)= 0.881 + 0.107\,$\theta ^2$
($\theta$ in arc minutes) as given in \citet{2005ApJSmurrayetal}, this leads to
a 90\% encircled energy radius of R$_{90}$=0$\farcs$91. Less than two photons
have been detected in the circular area spanned by R$_{90}$ (see
Figure~\ref{zoom1905}). From the number of photons detected in 300.8 ks we
calculate a 95\% confidence upper limit on the source count using both the
method explained in \citet{1986ApJ...303..336G} as well as the method explained
in the appendix of \citet{2007ApJ...657.1026W}. Even though the number of
photons falling inside this area is less than 2 we take 2 as the number of
detected photons in what follows, to account for the fact that we took the 90\%
encircled energy radius (not 100\%), and to account for the slight smearing that
will have occurred due to the alignment of the images introduced by the finite
accuracy at which the boresight correction can be determined. Following
\citet{2007ApJ...657.1026W} we calculate the probability that the background
contributes fewer than the 2 photons inside the 90\% encircled energy radius.
This probability is~effectively~0. 

From the method of \citet{1986ApJ...303..336G}, we derive a 95\% confidence upper
limit on the source count of 6.3 which given the exposure time corresponds to an
upper limit on the count rate of 2.1$\times 10^{-5}$ counts s$^{-1}$.  In the
method of \citet{2007ApJ...657.1026W} m$_T$ and m$_R$ are defined, corresponding to
the number of X--ray photons detected in the Target and Reference apertures of
measure $\Omega_T$ and $\Omega_R$, respectively. For $\Omega_T$ we have taken a
circular region with radius of 1\arcsec~centered on the source position, as above
this gives $\approx$2 counts in the detection region. For $\Omega_R$ we have taken
an annulus with inner and outer radius of 4\farcs92 and 39\farcs36, respectively.
We have excluded a weak source from this background area. The size of the areas
$\Omega_T$ and $\Omega_R$ is 3.14 arcsec$^2$ and 4790.94 arcsec$^2$, respectively
(m$_T$ and m$_R$ are 2 and 2750, respectively). The expectation value on the number
of source counts for a confidence level of 95\% is $<$5 counts which corresponds to
an upper limit on the count rate of 1.7$\times 10^{-5}$ counts s$^{-1}$. In the
remainder of the Paper we conservatively use the slightly higher value from the
\citet{1986ApJ...303..336G} method.

We have used {\sl W3PIMMS}\footnote{available at
http://heasarc.gsfc.nasa.gov/Tools/w3pimms.html} to estimate 95\% confidence
limits on the source flux and luminosity in different X--ray bands employing the
various models often found for neutron star soft X--ray transients (SXTs) in
quiescence. The obtained limits are listed in Table~\ref{limits}. We have
estimated the effective temperature at the surface of the neutron star,
T$_{eff}$, and as measured at infinity, T$_{eff}^\infty$, as follows: we have
used the flux limit obtained for a 0.2 keV black body model as limit for the
neutron star atmosphere model in {\sl xspec} version 11.3.2p (\citealt{ar1996};
\citealt{1996A&A...315..141Z}). Ideally, one would like to use a response and
auxiliary response matrix specific for the location of the source on the CCD.
However, for the combined observation this is not possible, hence, we have used
the standard response files for on--axis ACIS--S observations. Given the
distance of 10 kpc, assuming a  1.4~M$_\odot$ neutron star with radius of 10 km,
a pure Hydrogen, non--magnetic atmosphere, the limits on T$_{eff}$ and
T$_{eff}^\infty$ are 4.6$\times 10^5$ K and 3.5$\times 10^5$ K, respectively.

\begin{deluxetable}{cccc}
\tabletypesize{\scriptsize}
\tablecaption{Upper limits to the unabsorbed source flux\tablenotemark{a} and luminosity\tablenotemark{b}.\label{limits}}
\tablewidth{0pt}
\tablehead{
\colhead{Model} & \colhead{F${\rm _{0.5-10\,keV}}$ unabs.} & \colhead{F${\rm _{0.01-10\,keV}}$ unabs.} & \colhead{L${\rm _{0.5-10\,keV}}$} \\
\colhead{} & \colhead{erg cm$^{-2}$ s$^{-1}$} & \colhead{erg cm$^{-2}$ s$^{-1}$} & \colhead{${\rm (\frac{d}{10~kpc})^2}$ erg s$^{-1}$}\\
}
\startdata
PL\tablenotemark{c} $\Gamma$=2.0 &  2.0$\times 10^{-16}$& 4.5$\times 10^{-16}$ & 2.4$\times 10^{30}$\\
BB\tablenotemark{c} T=0.2 keV&  1.4$\times 10^{-16}$& 1.9$\times 10^{-16}$  & 1.6$\times 10^{30}$\\
BB\tablenotemark{c} T=0.1 keV&  2.2$\times 10^{-16}$& 8.8$\times 10^{-16}$ & 2.6$\times 10^{30}$\\

\enddata
\tablenotetext{a}{Unabsorbed flux (F) is given in the 0.5--10 keV and 0.01--10 keV band. The used interstellar extinction is 2.1$\times 10^{21}$ cm$^{-2}$. }
\tablenotetext{b}{The luminosity (L) is given for 0.5--10 keV and a distance of 10 kpc.}
\tablenotetext{c}{PL stands for power law and BB for blackbody.}

\end{deluxetable}

\begin{figure} \epsscale{.80} \plotone{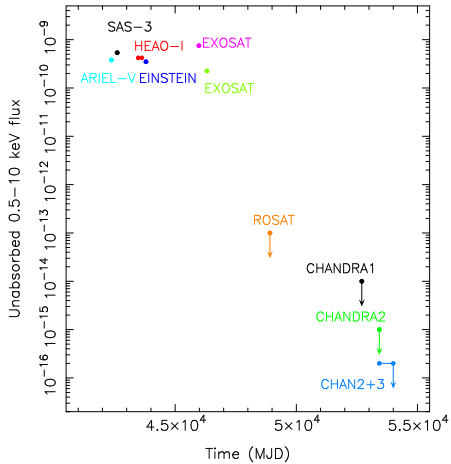} \caption{The unabsorbed 0.5--10
keV X--ray flux history of \src. The range in dates plotted for the HEAO--I
satellite indicates the time span over which the source was detected on multiple
occasions by that satellite. The upper limit indicated with "CHAN2+3" has been
derived using {\it Chandra} data from 2005 and 2006 (see Table~\ref{log}). In
order to convert fluxes given in the literature in other X--ray bands to the
0.5--10 keV band we assume the spectrum to be well--represented by an absorbed
power law with index 2 (except for the ROSAT and {\it Chandra} upper limit which
were derived assuming a 0.3 and 0.2 keV black body, respectively). Arrows on the
data points indicate upper limits to the X--ray flux.\label{flux}} \end{figure}

Using the stringent limit on the source luminosity together with detections and
limits on the detection of the source obtained at other times we have
constructed the long term X--ray lightcurve of \src\ (see
Fig.~\ref{flux})\footnote{Data for Fig.~\ref{flux} was taken or derived from
\citet{1976MNRAS.175P..39S} (Ariel--V), \citet{1976MNRAS.177P..93L} (SAS--3),
\citet{1980AJ.....85.1062R} (HEAO--I), \citet{1997ApJS..109..177C} (Einstein),
\citet{1990A&A...228..115C} (EXOSAT, MJD~46316), \citet{1995A&AS..109....9G}
(EXOSAT, MJD~45982),  \citet{1999A&Avogesetal} (ROSAT),
\citet{2005ApJ...627..926J} (Chandra1),  \citet{2006MNRAS.368.1803J} (Chandra2),
and this work (Chandra 2+3)}. From Fig.~\ref{flux} it is clear that the source
underwent an outburst with a duration of at least 9.9 years (the time between
the first and last detection). Using the SAS--3 satellite several type I X--ray
bursts were detected (\citealt{1976IAUC.2983....1L} [multiple bursts],
\citealt{1976MNRAS.177P..93L} [5], \citealt{1976IAUC.2984....2L} [1]). However,
since the persistent flux level was in most cases not quoted only 1 SAS--3 point
appears in Fig.~\ref{flux}. 

\subsection{Magellan optical observations}

We obtained Sloan $i'$--band images using the Magellan Instant Camera (MagIC)
instrument mounted on the 6.5~m Magellan--Clay telescope at Las Campanas
Observatory. On June 23, 2006 (MJD 53909 UTC), three 300 second exposures were
collected between 5:02--5:19 UTC. The observing conditions were excellent with a
photometric sky and a seeing of 0\farcs4. MagIC delivers a 2.35\arcmin\ field of
view sampled at 0\farcs069/pixel. Frames were readout in quad amplifier mode, and
were debiased and then flatfielded using dithered twilight sky observations.

We astrometrically calibrated the median combined CCD image by matching the
positions of stars in the image against those of stars from images presented in
Jonker et al.~(2006). The latter images were calibrated with respect to the second
version of the USNO CCD Astrograph Catalog (\citealt{2004AJ....127.3043Z})
to provide a ICRS J2000.0 astrometric frame.

Photometric calibration was performed using images obtained on June 25, 2006
with the 4-m Blanco telescope and its MOSAIC imager at CTIO. A 120s Sloan
$i'$--band image of the field surrounding \src\ was obtained under photometric
conditions together with observations of the standard star fields PG~1323 and
SA~107. These allowed us  to perform an absolute magnitude calibration of stars
in the proximity of our target. Since we only observed the field of \src\ in one
filter no color terms have been calculated. 

\begin{figure}
\epsscale{.80}
\plotone{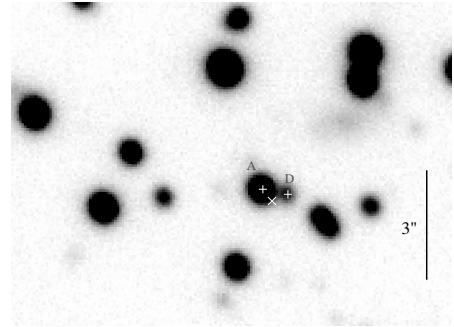}
\caption{Deep, 3$\times$5 minutes integration, sloan $i$ filter Magellan
observation  of the field of \src\ obtained with the MagIC instrument under a
seeing of 0\farcs4. The cross indicates the outburst source position. The "+" signs
labelled A and D are nearby field stars (labels follow those in Jonker et
al.~2006). The vertical bar has a length of 3\arcsec. North is up and East is left.
\label{sloani}}
\end{figure}

Despite the excellent observing conditions at Magellan, the counterpart of \src\ in
quiescence is not detected down to a 5$\sigma$ magnitude limit of $i>25.3$.  A high
resolution finder chart is presented in Fig.~\ref{sloani}.

\section{Discussion}

We have obtained a 300 ks--long {\it Chandra} observation of the field of \src.
Using this observation we do not detect the quiescent X--ray counterpart to this
neutron star soft X--ray transient. The limit on the source flux depends on the
assumed spectral model. For instance, a spectral model of an absorbed power law
with index of 2 as observed for weak quiescent neutron star transients and
quiescent black hole X-ray transients (\citealt{2004MNRAS.354..666J},
\citealt{2002ApJ...570..277K}), gives a 95\% confidence upper limit on the
0.5--10 keV source flux of 2.0$\times 10^{-16}$ erg cm$^{-2}$ s$^{-1}$. Given
the upper limit on the source distance of 10 kpc (\citealt{1990A&A...228..115C},
\citealt{2004MNRAS.354..355J}) this converts to a 0.5--10 keV luminosity limit
of $2.4\times 10^{30} {\rm\,erg\,s^{-1}}$. This means that the luminosity of
this neutron star SXT is lower than that observed for the weakest black hole SXT
A~0620--00 which has for a distance of 1 kpc an unabsorbed quiescent 0.5--10 keV
luminosity of $3\times 10^{30}$ erg s$^{-1}$ (\citealt{2001ApJ...553L..47G},
\citealt{2002ApJ...570..277K}). Note that some black hole SXTs have not been
detected in quiescence, and whereas the present upper limit to the luminosity in
those cases is well above that derived for \src\ one cannot exclude that deeper
observations will reveal a lower luminosity than that derived for \src.
Nevertheless, from present data the claim that there is evidence for a black
hole event horizon from a lower quiescent luminosity in black holes than neutron
stars (e.g.~\citealt{2001ApJ...553L..47G}) is unproven and at least does not
hold universally.

Scaling of the observed quiescent luminosity with the Eddington luminosity in order
to try to normalize the neutron star and black hole systems to the same mass
accretion rate is very uncertain. For instance, since an unknown amount of the
transfered mass might be lost from the system in the form of a disk wind
(e.g.~\citealt{2006Natur.441..953M}), the relation between the mass transfer and
mass accretion rate is not well constrained. This might be especially important for
neutron star systems in quiescence if a propellor regime exists
(\citealt{1975A&A....39..185I}). On the other hand, as shown for instance by
\citet{2001MNRAS.322...31F}, \citet{2003MNRAS.343L..99F} and
\citet{2006MNRAS.370.1351G}, black holes are producing more powerful jets
observable in radio. Matter could also be lost from these systems via these jets,
the amount depends on the unknown composition of the jets. Finally, it is unclear
whether accretion in low--tranfer states proceeds via an advection dominated
accretion flow (e.g.~ADAF; \citealt{1995ApJ...452..710N}) or via a disk
(e.g.~\citealt{2003ApJ...593..184L}). 

In deep optical Sloan $i$--band images obtained with the 6.5~m Magellan telescope
under excellent conditions (seeing 0\farcs4) we do not detect the optical
counterpart to \src. For a distance of 10 kpc the magnitude upper limit of $i>$25.3
converts into an upper limit on the absolute magnitude of $I=$9.6 (as in
\citealt{2006MNRAS.368.1803J} we converted the observed outburst N${\rm _H}$ from
\citealt{1997ApJS..109..177C} to a reddening A$_i=0.7$ in the sloan $i$ band using
the conversion factors given in \citealt{1985ApJ...288..618R} and
\citealt{scfida1998}). This limit implies that the companion star of \src\ has a
spectral type later than M5 or is a white dwarf as in ultra--compact X--ray
binaries. It strengthens the identification of \src\ as an ultra--compact X--ray
binary (\citealt{2006MNRAS.368.1803J}).

An absorbed thermal neutron star atmosphere spectral model for a distance of 10
kpc gives a limit to the effective temperature at the surface of the neutron
star of 4.6$\times 10^5$ K (for a neutron star mass and radius of 1.4 M$_\odot$
and 10 km, respectively and assuming a pure Hydrogen non--magnetic atmosphere).
For such a neutron star this implies an effective temperature of 3.5$\times
10^5$ K at infinity.

A factor in determining the limit on the source flux is the amount of interstellar
extinction that is assumed. We have taken the conservative value for the Hydrogen
column density of ${\rm N_H =2.1\times 10^{21}\,cm^{-2}}$ from the results of
\citet{1997ApJS..109..177C} who found ${\rm N_H =(1.9\pm0.2) \times
10^{21}\,cm^{-2}}$. Those authors found that the Hydrogen column densities derived
for several LMXBs from their spectral fits to Einstein data agree with values found
using ROSAT spectra. Besides and related to the N${\rm _H}$, the bolometric
correction is important in the conversion of the 0.5--10 keV band limit to a
bolometric luminosity limit. E.g.~from the upper limit on the effective temperature
of a neutron star atmosphere an upper limit to the bolometric luminosity of
$10^{31}$ erg s$^{-1}$ is determined implying a bolometric correction of
$\approx$4--5. For a power law spectral model with index 2, the bolometric
correction would be $\approx$3 (here we have taken the 0.01--100 keV luminosity as
a good measure of the bolometric luminosity). In this we again have taken ${\rm N_H
=2.1\times 10^{21}\,cm^{-2}}$.

The new deep limit on the quiescent thermal X--ray emission of \src\ implies that
the neutron star must cool faster than possible with modified URCA processes even
for time averaged mass accretion rates as low as 10$^{-13}$ M$_\odot$ year$^{-1}$
(\citealt{2006MNRAS.368.1803J}), unless the neutron star core is not in a steady
state. This would mean that \src\ had been in quiescence prior to the $\sim$10 year
long outburst for $\approx$10 thousand years since the core reaches steady state on
such timescales (\citealt{2001ApJ...548L.175C}, \citealt{2004ARA&A..42..169Y}). On
the other hand, here again one should keep in mind that the relation between the
mass transfer and mass accretion rate is not well known especially for quiescent
systems. So, it is possible that the time averaged mass transfer rate is larger
than 10$^{-13}$ M$_\odot$ year$^{-1}$ whereas the mass accretion rate onto the
neutron star is lower than this.

\acknowledgments

PGJ acknowledges Ed Brown for useful discussions, support from NASA
grant G06-7032X and support from the Netherlands Organisation for Scientific
Research. DS acknowledges support through the NASA Guest Observer program as
well as a PPARC/STFC Advanced Fellowship. This paper includes data gathered with
the 6.5~m Magellan Telescopes located at Las Campanas Observatory, Chile. {\it
Facilities:} \facility{Magellan}, \facility{CXO (ACIS)}.

\end{document}